\documentclass[prl,amsmath,amssymb,twocolumn,superscriptaddress,showpacs]{revtex4-1}
\usepackage{bm}
\usepackage{amssymb}
\usepackage{colordvi}
\usepackage{graphicx}
\usepackage{color}
\usepackage{hyperref}
\usepackage{ulem}

\newcommand{\beq}{\begin{equation}}
\newcommand{\eeq}{\end{equation}}
\newcommand{\bea}{\begin{eqnarray}}
\newcommand{\eea}{\end{eqnarray}}

\begin{document}

\title{Higher-Order Weyl Semimetals}

\affiliation{School of Physical Science and Technology, \& Collaborative Innovation Center of Suzhou Nano Science and Technology, Soochow University, 1 Shizi Street, Suzhou 215006, China}
\affiliation{School of Physical Science and Technology, Guangxi Normal University, Guilin 541004, China}
\affiliation{Department of Physics and Center for Theoretical Sciences, National Taiwan University, Taipei 10617, Taiwan}
\affiliation{Physics Division, National Center for Theoretical Sciences, Hsinchu 30013, Taiwan}

\author{Hai-Xiao Wang}
\affiliation{School of Physical Science and Technology, \& Collaborative Innovation Center of Suzhou Nano Science and Technology, Soochow University, 1 Shizi Street, Suzhou 215006, China}
\affiliation{School of Physical Science and Technology, Guangxi Normal University, Guilin 541004, China}
\affiliation{Department of Physics and Center for Theoretical Sciences, National Taiwan University, Taipei 10617, Taiwan}
\affiliation{Physics Division, National Center for Theoretical Sciences, Hsinchu 30013, Taiwan}

\author{Zhi-Kang Lin}
\affiliation{School of Physical Science and Technology, \& Collaborative Innovation Center of Suzhou Nano Science and Technology, Soochow University, 1 Shizi Street, Suzhou 215006, China}

\author{Bin Jiang}
\affiliation{School of Physical Science and Technology, \& Collaborative Innovation Center of Suzhou Nano Science and Technology, Soochow University, 1 Shizi Street, Suzhou 215006, China}

\author{Guang-Yu Guo}
\affiliation{Department of Physics and Center for Theoretical Sciences, National Taiwan University, Taipei 10617, Taiwan}
\affiliation{Physics Division, National Center for Theoretical Sciences, Hsinchu 30013, Taiwan}

\author{Jian-Hua Jiang}
\email{jianhuajiang@suda.edu.cn}
\affiliation{School of Physical Science and Technology, \& Collaborative Innovation Center of Suzhou Nano Science and Technology, Soochow University, 1 Shizi Street, Suzhou 215006, China}

\date{\today}

\begin{abstract}
	Higher-order topology yields intriguing multidimensional topological phenomena, while Weyl semimetals have unconventional properties such as chiral anomaly. However, so far, Weyl physics remain disconnected with higher-order topology. Here, we report the theoretical discovery of higher-order Weyl points and thereby the establishment of such an important connection. We demonstrate that higher-order Weyl points can emerge in chiral materials such as chiral tetragonal crystals as the intermediate phase between the conventional Weyl semimetal and 3D higher-order topological phases. Higher-order Weyl semimetals manifest themselves uniquely by exhibiting concurrent chiral Fermi-arc surface states, topological hinge states, and the momentum-dependent fractional hinge charge, revealing a novel class of higher-order topological phases. 
\end{abstract}

\maketitle

{\it Introduction.---} Topological insulators and semimetals with bulk-edge correspondence have invoked paradigm shifts in the study of condensed matter physics~\cite{Hasan2010,Qi2011}, photonics~\cite{Ozawa2019}, acoustics~\cite{Zhangxj2018,Ma2019} and phononics~\cite{Liu2019}. Recent discovery of higher-order topological insulators (HOTIs)~\cite{Hughes2017Sci,Hughes2017prb,Langbehn2017,Song2017,Schindler2018,Huber2018,Bahl2018,Imhof2018,Noh2018,Ezawa2018,AQTI,Hafezi2019,Bernevig2020qudrupole1,Zhenbo2019quadrupole2,Xuyong2019qudrupole3,JJH2020quadrupole,JJH2020polariton,Meng2020quadrupole,LuMH2018dipole,Zhangbl2018Kagome,Khanikaev2018kagome,JJH2019Natphys,Hassan2019,Zhangshuang2019,Christensen2019,Xiabz2019elastic,Iwamoto2019dipole,DongJW2019dipole,LuMH2019dipole,JJH2020surfacewave} which exhibit gapped edge states and topologically-protected corner/hinge states, unveils a new horizon beyond the conventional bulk-edge correspondence. The underlying bulk-edge-corner/hinge correspondence manifests  concurrent multidimensional topological physics, which is the most salient feature of higher-order topology. On the other hand, Weyl points have been one of the focuses in the study of topological physics and materials, due to their anomalous physical properties (e.g., chiral anomaly) and their connection with various topological phases (e.g., three-dimensional (3D) quantum Hall effects)~\cite{Wan2011Weyl,JJH2012,Lu2013,Xiao2015,Lu2015,Fang2016helicoid,Yan2017AnnualRev,Lifeng2017,Gehao2018,Zhangs2018idealWeyl,Mele2018}. However, so far, Weyl points have not yet been been connected with higher-order topology.

\begin{figure}
	\includegraphics[width=3.4in]{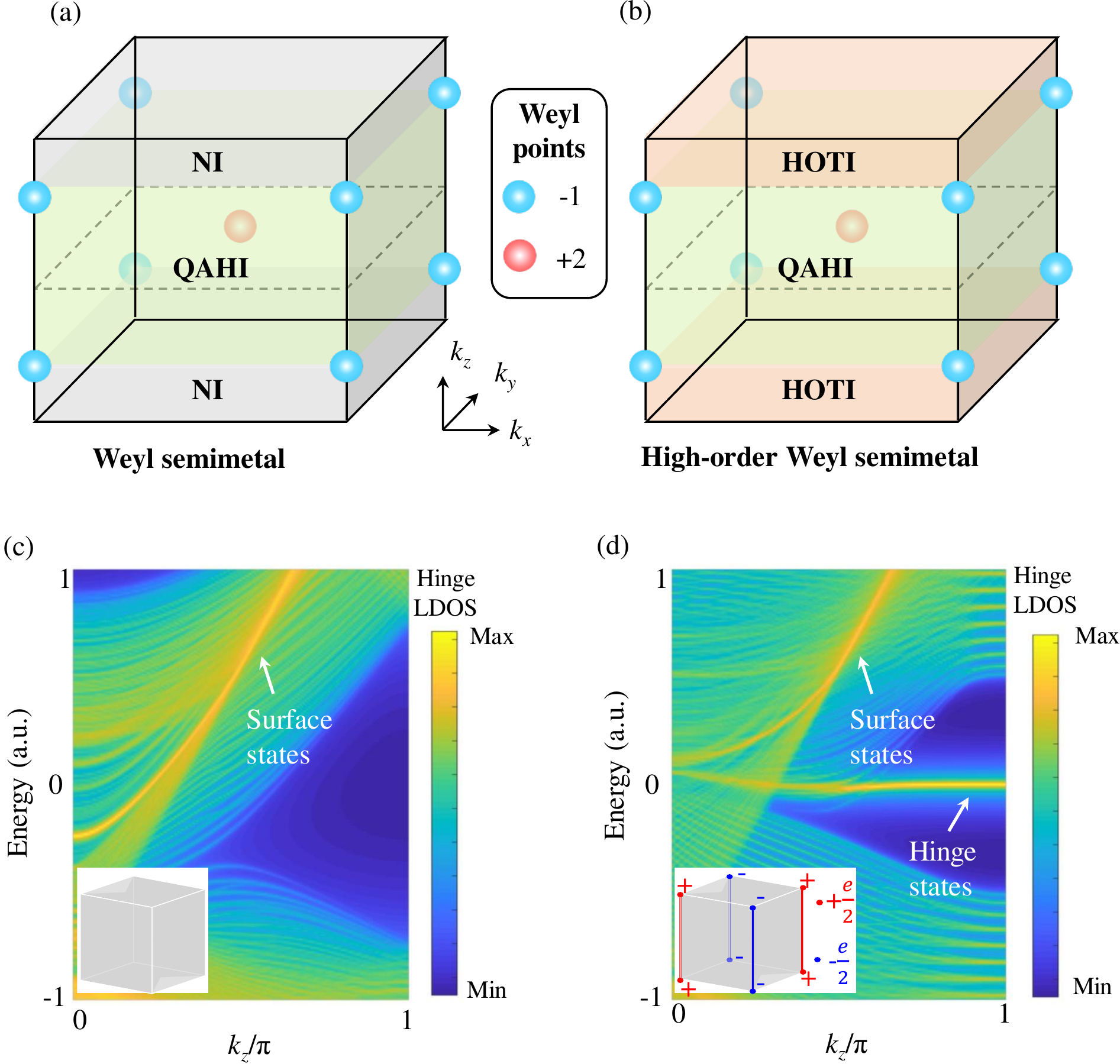}
	\caption{(Color online) (a)-(b): Schematic of (a) conventional and (b) higher-order Weyl semimetals. Weyl points and their topological charges are depicted in the middle. (c)-(d): Hinge LDOS for (c) conventional and (d) higher-order Weyl semimetals. Higher-order Weyl semimetal exhibits concurrent hinge and surface states and fractional hinge charges (see insets) in finite systems. Finite structures with 20$\times$20 unit-cells in the $x$-$y$ plane are adopted in the calculation using models in Fig.~2.}
\end{figure}

In this Letter, we establish such a missing connection. The basic picture is illustrated in Fig.~1. If a 3D system is regarded as $k_z$ dependent two-dimensional (2D) systems, conventional Weyl points separate the 3D Brillouin zone into regions with quantum anomalous Hall insulator (QAHI) and normal insulator (NI) phases [Fig.~1(a)]~\cite{JJH2012,Xiao2015,Lifeng2017,Gehao2018}. In contrast, higher-order Weyl points (HOWPs) separate the 3D Brillouin zone into regions with QAHI and HOTI phases [Fig.~1(b)]. As a consequence, when measuring the hinge local density-of-states (LDOS), conventional Weyl semimetals exhibit only spectral features of topological surface states [Fig.~1(c)]~\cite{JJH2012,Lu2015,Fang2016helicoid,Lifeng2017,Gehao2018,Zhangs2018idealWeyl}. In comparison, the higher-order Weyl semimetals exhibit both spectral features of the surface states and the higher-order hinge states [Fig.~1(d)]. In addition, due to the higher-order topology, fractional charge can emerge on the hinges of higher-order Weyl semimetals. The concurrent chiral Fermi-arc surface states, topological hinge states, and the momentum-dependent fractional hinge charge emerge as the hallmark characteristics of higher-order Weyl semimetals.

{\it Model for the HOWPs.}---To realize the HOWPs depicted in Fig.~1, we first notice that both QAHIs and quadrupole topological insulators (QTIs) can be realized using synthetic gauge fluxes~\cite{Haldane1988,JJH2012,Xiao2015,Bahl2018,Hafezi2019}. QAHI is a ``first-order'' 2D topological insulator with quantized Hall conductance and chiral edge states~\cite{Haldane1988}. In comparison, QTI is a ``second-order'' 2D topological insulator that hosts gapped edge states, topological corner states and fractional corner charge $\pm 1/2$ with a quadrupole configuration~\cite{Hughes2017Sci,Hughes2017prb}. We propose a tight-binding model with square-spiral structures where translation along the $z$ direction generates synthetic fluxes straightforwardly.

The tight-binding model based on tetragonal-lattices is depicted in Fig.~2(a). For simplicity, the lattice constants along the $x,y,z$ directions are set to unity. In each unit-cell, there are four identical sites (indicated by the black spheres) arranged with equal spacing along the $x$, $y$ and $z$ directions. The structure exhibits square-spirals as going from 1$\to$3$\to$2$\to$4, yielding a fourfold screw symmetry $S_{4z}:= (x,y,z) \rightarrow (y,-x,z+\frac{1}{4})$. In addition, there are twofold rotation symmetries $C_{2x}:=(x,y,z) \rightarrow (x,-y,-z)$ and $C_{2y}:=(x,y,z) \rightarrow (-x,y,-z-\frac{1}{2})$ which, together with the time-reversal symmetry, protect the quadrupole topology.

\begin{figure}[hbtp]
	\includegraphics[width=3.4in]{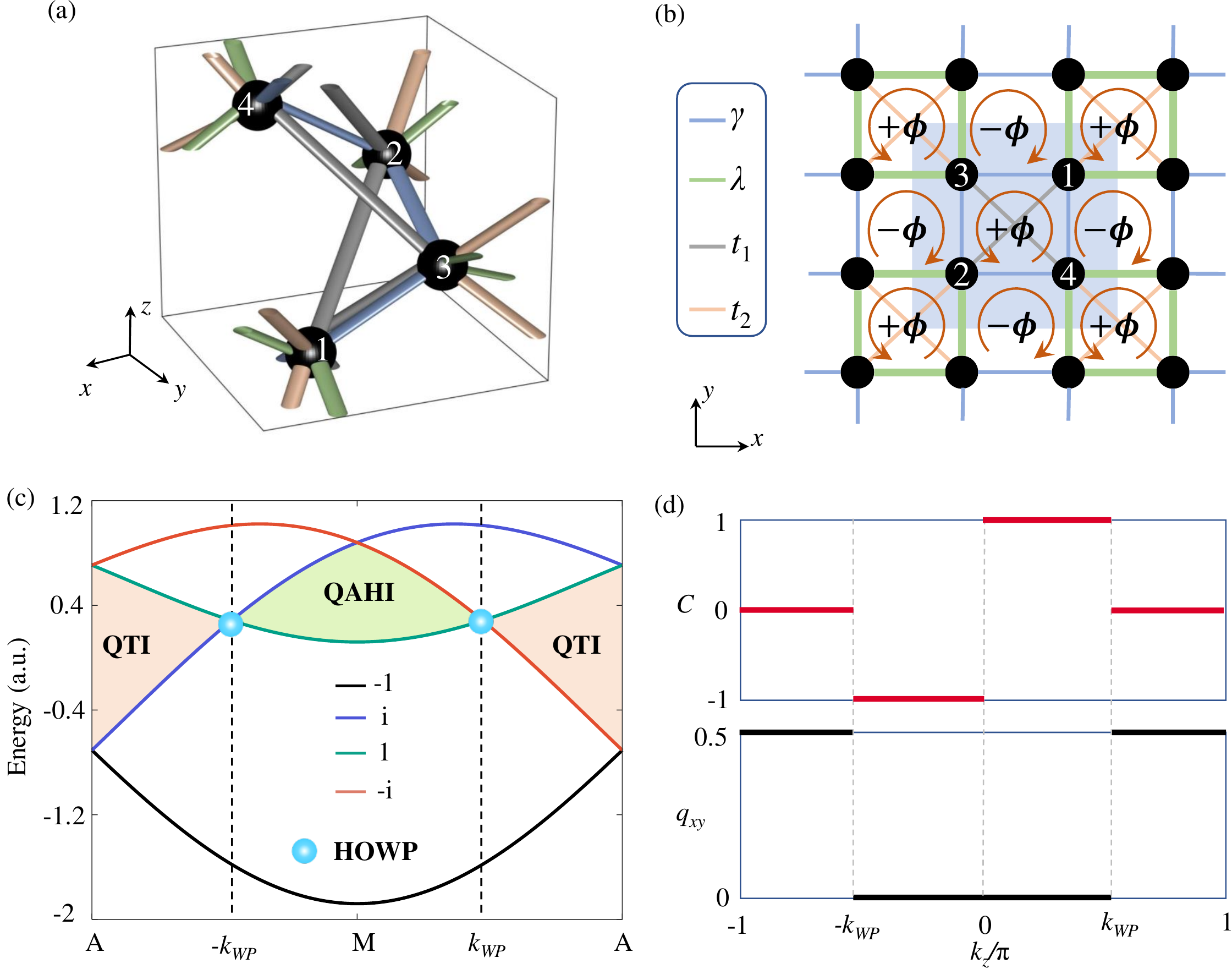}
	\caption{(a) 3D tight-binding model with chiral couplings in tetragonal lattices. (b) Top-down view of the model. Brown arrows represent the spiral hopping structures which induce the synthetic gauge fluxes $\pm \phi$. (c) Evolution of energy bands and $\tilde{C}_{4z}$ eigenvalues along the A-M-A line. The gap between the second and the third bands experiences topological transitions from QTI to QAHI via the HOWPs. (d) Evolution of Chern number ${\cal C}$ (upper) and quadrupole index $q_{xy}$ (lower) with $k_z$ for the gap between the second and the third bands. Parameters for higher-order Weyl phase: $\gamma=0.6$, $\lambda=1$, $t_1=0.2$ and $t_2=0.24$. The conventional Weyl phase in Fig.~1 is realized by the following exchanges: $\lambda\leftrightarrow \gamma$, $t_1\leftrightarrow t_2$.}
\end{figure}

The tight-binding Hamiltonian is given by
\beq
\begin{split}
	H=\sum_{<i,j>} t_{NN}(c_{i}^\dagger a_{j}+d_{i}^\dagger b_{j}+c_{i}^\dagger b_{j} +a_{i}^\dagger d_{j})\\ 
	+\sum_{<<i,j>>}t_{NNN} (a_i^\dagger b_j+c_i^\dagger d_j)+h.c.,
\end{split}
\eeq
where $a(a^\dagger), b(b^\dagger), c(c^\dagger), d(d^\dagger)$ are the annihilation (creation) operators on the sites 1,2,3,4, separately. In the right side of the above equation, the first term describes the nearest-neighbor couplings, where $t_{NN}=\gamma$ ($\lambda$) for the intra-(inter-) unit-cell coupling. The second term represents the next-nearest-neighbor couplings, where $t_{NNN}=t_1$ ($t_2$) for the intra-(inter-) unit-cell coupling.

A top-down view is shown in Fig.~2(b) where the spiral directions are labeled by the arrows for different regions. The spiral structures introduce synthetic gauge fluxes $\pm \phi$ into these regions. Remarkably, such synthetic gauge fluxes are merely due to the translation along the $z$ direction which can be understood as follows: The site 4 in a unit-cell of central coordinate $(X, Y, Z)$ couples only to the site 1 in the upper unit-cell of central coordinate $(X, Y, Z+1)$. Such a coupling naturally picks up a translational phase $\phi=k_z$. We remark that our model, with time-reversal symmetry and using only positive couplings, is more feasible than other models for quadrupole topology where either a mixture of positive and negative couplings or magnetic fluxes are required~\cite{Hughes2017Sci,Hughes2018semimetal,AQTI}.

The synthetic gauge fluxes are crucial in our construction. First, they give rise to the emergence of the QAHIs with Chern number ${\cal C}=1$ (-1) for $0<k_z<k_{WP}$ ($-k_{WP}<k_z<0$). Second, they yield the QTI phase with a quadrupole index $q_{xy}=\frac{1}{2}$ for $|k_z|>k_{WP}$ (see Supplemental Materials for the characterizations of these two topological phases~\cite{SM}). Here, the HOWPs appear at the wavevectors $(\pi, \pi, \pm k_{WP})$ with a topological charge of -1. According to the Nielsen-Ninomiya theorem~\cite{Nielsen1981}, to compensate the nonzero total charge of these HOWPs~\cite{Nielsen1981}, a quadratic Weyl point with charge $+2$ appears at the $\Gamma$ point. If the 3D system is viewed as $k_z$-dependent 2D systems, the HOWPs can then be regarded as the transitions between the QAHIs and QTIs.

The Chern number and the quadrupole index can be determined through two approaches. First, straightforward calculations based on the Wilson-loop and nested Wilson-loop approaches confirm the above picture (see Supplemental Material for details~\cite{SM}). The eigenvalues of the (nested) Wilson-loop operator gives the (nested) Wannier bands. The region with $|k_z|<|k_{WP}|$ exhibits gapless Wannier bands of which the winding property gives the Chern numbers shown in Fig.~2(d). The change of the Chern number ${\cal C}$ at $k_z=0$ is due to the quadratic Weyl point at the $\Gamma$ point, while the change at $k_z=\pm k_{WP}$ is due to the HOWPs. In comparison, the region with $|k_z|>|k_{WP}|$ exhibits gapped Wannier bands. The nested Wannier bands yield quantized Wannier band polarizations and nontrivial quadrupole index $q_{xy}=\frac{1}{2}$ (see Supplemental Material for details~\cite{SM}). The quantization of the Wannier band polarizations and the quadrupole index to 0 or $\frac{1}{2}$ are dictated by the symmetry operators, $\Theta_x=C_{2x} {\cal T}$ and $\Theta_y=C_{2y} {\cal T}$, where ${\cal T}$ is the time-reversal operator (see proof in the Supplemental Material~\cite{SM}). The noncommutativeness of the $\Theta_x$ and $\Theta_y$ operators are crucial for the emergence of the gapped Wannier bands~\cite{Hughes2017prb}. Though the synthetic flux model is quite different from the original model for quadrupole topological insulators proposed in Ref.~\cite{Hughes2017Sci}, it does support quadrupole topological band gap with different protective symmetries as validated in details in the Supplemental Material~\cite{SM}.

On the other hand, the quadrupole index can be extracted from the rotation eigenvalues at the high-symmetry points~\cite{Hughes2017prb,Zhenbo2019quadrupole2}. Specifically, for each $k_z$, we examine the eigenvalues of the pseudo-rotation operator $\tilde{C}_{4z}=e^{-ik_z/4}S_{4z}$ at the $\tilde{\Gamma}=(0,0,k_z)$ and $\tilde{M}=(\pi,\pi,k_z)$ points. The first band always has $\tilde{C}_{4z}(\tilde{\Gamma})=1$, while the second band has $\tilde{C}_{4z}(\tilde{\Gamma})=i$ ($-i$) for $k_z>0$ ($k_z<0$). In comparison, the rotation eigenvalues at the $\tilde{M}$ point vary with $k_z$, as presented in Fig.~2(c). The quadrupole index is related to the rotation eigenvalues as~\cite{Hughes2017prb,Zhenbo2019quadrupole2}
\begin{align}
& \exp{(i2\pi q_{xy})}=r_4^-(\tilde{\Gamma})r_4^-(\tilde{M})^\ast.
\end{align} 
Here, $r_4^-$ denotes the $\tilde{C}_{4z}$ eigenvalues of $\pm i$. From the $\tilde{C}_{4z}$ eigenvalues in Fig.~2(c), one obtains quadrupole topological numbers consistent with those in Fig.~2(d). The band inversion between the second and third bands along the A-M-A line triggers the transition between the QAHI phases and the QTI phase, and thus leads to the formation of the HOWPs.

{\it Topological surface states}.---The coexistence of the conventional and higher-order topology is manifested in the topological surface states. Depending on the wavevector $k_z$, the topological surface states can be gapless or gapped. As shown in Fig.~3(a), the surface states are gapless along the $\bar{\Gamma}\bar{Y}$ line, while the surface states become gapped along the $\bar{R}\bar{Z}$ line. The former originates from the gapless chiral edge states in the QAHI phases, while the latter results from the gapped edge states in the QTI phase.

We further study the topological surface states by examining the surface LDOS at given energies. We sample three different energies: $E_1$ is exactly at the energy of the quadratic Weyl points, $E_2$ is exactly at the energy of the HOWPs, $E_3$ is an energy above. At $E_1$ and $E_2$, there are Fermi-arc surface states connecting the projection of the quadratic Weyl point at $\bar{\Gamma}$ and the projections of the HOWPs at the surface Brillouin zone boundary. These chiral Fermi arcs connecting projections of Weyl points with opposite charges are consistent with the conventional picture~\cite{Fang2016helicoid}. Nevertheless, the iso-energy curves form noncontractible loops winding around the torus of surface Brillouin zone, which is different from the short Fermi arcs observed in many electronic systems~\cite{Mele2018}. In comparison, at $E_3$, the contractible Fermi arcs emerging at large $|k_z|$, which do not connect Weyl points with opposite topological charges, originate from the gapped surface states due to the quadrupole topology.

\begin{figure}
	\includegraphics[width=3.4in]{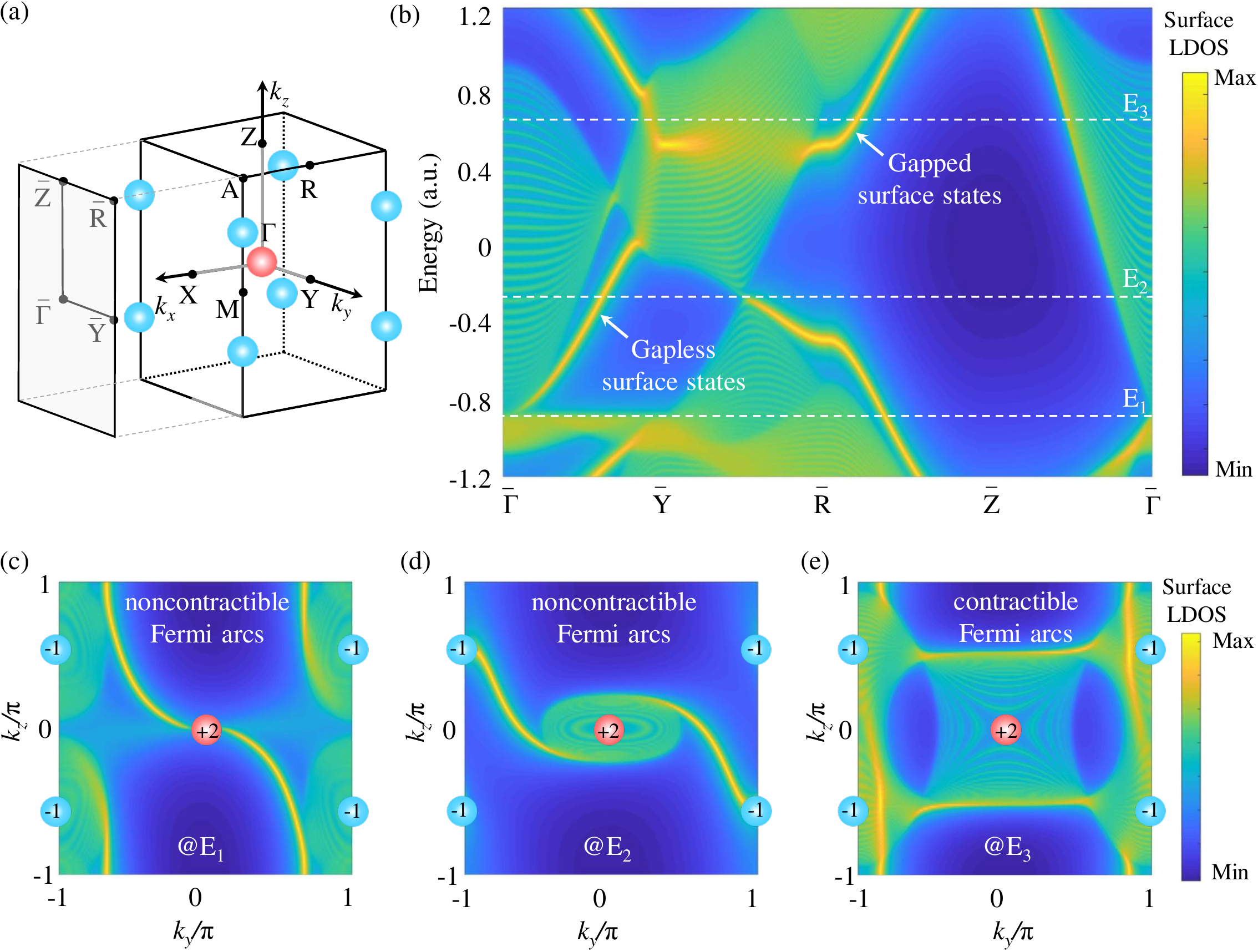}
	\caption{(a) Bulk and (100) surface Brillouin zones. (b) Surface LDOS for the (100) surface along high-symmetry lines. (c)-(e): Surface LDOS at three different energies labeled in (b) as dashed lines. $E_1=-0.85$ ($E_2=-0.27$) is at the energy of the quadratic Weyl points (the HOWPs), while $E_3=0.6$.}
\end{figure}

{\it Topological hinge states and fractional hinge charge}.---We now turn to the manifestation of the higher-order topology at the hinges. First, taking $k_z=\pi$ as an example to illustrate the underlying physics, the bulk and surface spectra are depicted in Fig.~4(a), which shows clearly the gapped surface states within the bulk band gap. For a finite system with four hinges, the eigen-spectrum exhibits four hinge states within the surface band gap whose wavefunctions are localized around the four hinges. The concurrent bulk, surface and hinge states appearing for $|k_z|>k_{WP}$ [as shown in Fig.~1(d)] demonstrate straightforwardly the higher-order topology.

One of the remarkable properties of the QTI is the fractional corner charge~\cite{Hughes2017Sci,Hughes2019corcharge}, $1/2$, with a quadrupole configuration as depicted in the inset of Fig.~4(c). We calculate the hinge charge by filling up to the first two bulk bands through $Q_h=\sum_{\alpha}P_\alpha^{hinge}$ where $P_\alpha^{hinge}$ is the probability of finding an electron in the hinge region for a filled state $\alpha$. Using this method, we obtain the fractional hinge charge $Q_h$ for various $k_z$ by considering a hinge region with $4\times 4$ unit-cells in a finite system consisting of $20 \times 20$ unit-cells in the $x$-$y$ plane (see Supplemental Material for details~\cite{SM}). It is seen from Fig.~4(c) that the hinge charge gradually reaches to the quantized value of 1/2 for $|k_z|>k_{WP}$. The crossover behavior around the Weyl point is due to the finite-size effects.

\begin{figure}
	\includegraphics[width=3.4in]{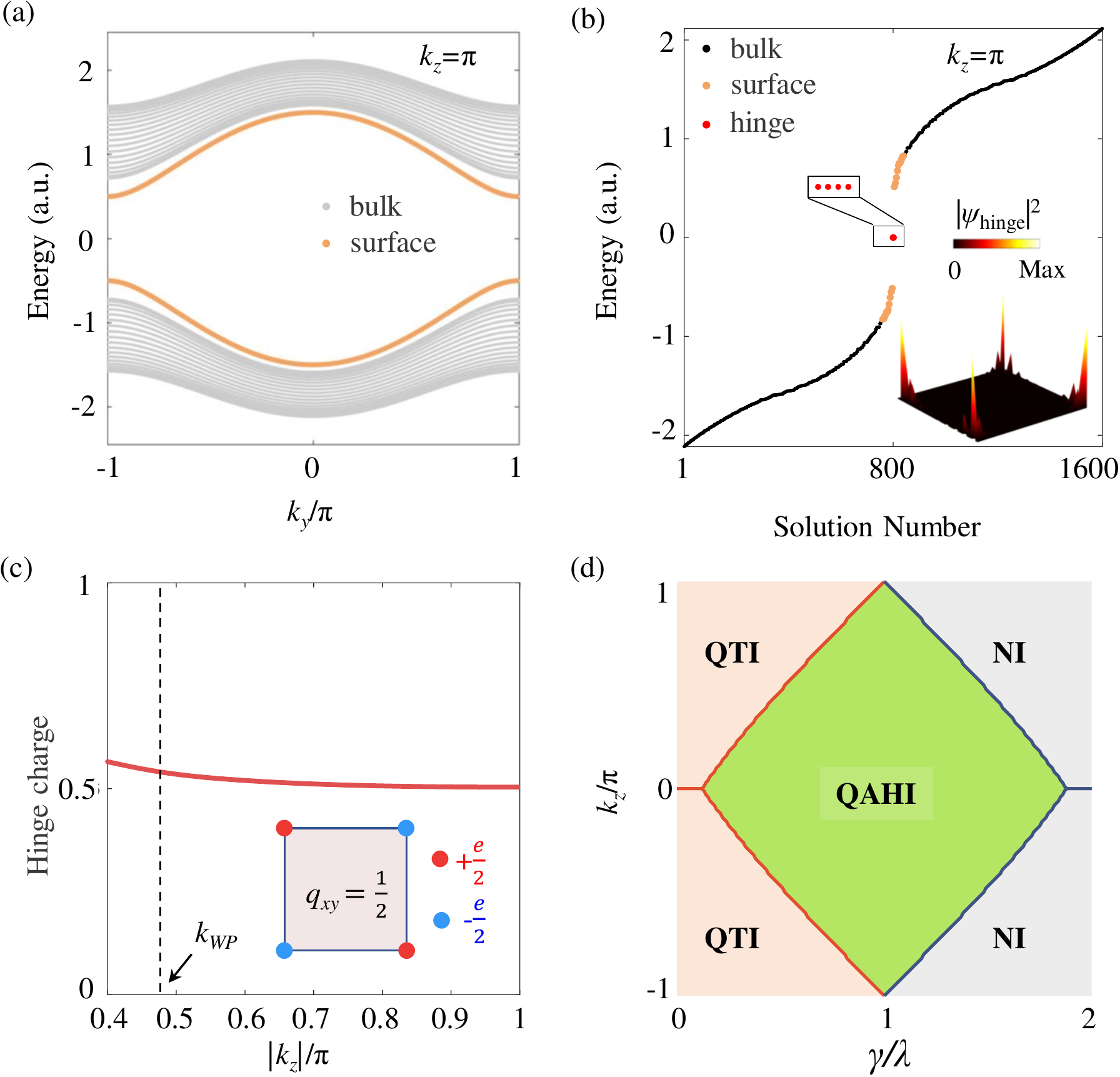}
	\caption{(a) Bulk and (100) surface spectra for $k_z=\pi$, showing gapped surface states. (b) Eigen-spectrum for a finite-sized system with $40 \times 40$ unit-cells in the $x$-$y$ plane for $k_z=\pi$, showing coexisting gapped bulk and surface states together with the fourfold degenerate hinge states. Inset shows the wavefunction of the hinge states. (c) Fractional hinge charge vs. $k_z$ calculated for a finite system with $40 \times 40$ unit cells in the $x$-$y$ plane. Inset shows the quadrupole configuration of the hinge charge. (d) Phase diagram of the system with $\lambda=1$, $t_1=0.2$ and $t_2=0.24$.}
\end{figure}

{\it Phase diagram.}---Finally, we study the phase diagram when the intra-unit-cell nearest-neighbor coupling $\gamma$ is varied. In Fig.~4(d) we present the evolution of the Weyl points along the A-M-A line with the parameter $\gamma$. It is seen that for $\gamma<\lambda$, the system is a higher-order Weyl semimetal. $k_{WP}$ decreases with decreasing $\gamma$. In the regime $\gamma<0.12\lambda$, we find that $k_{WP}\to 0$ and the system becomes a 3D quadrupole topological phase. In comparison, the system is a conventional Weyl semimetal when $\gamma>\lambda$. Therefore, the higher-order Weyl semimetal phase emerges as the intermediate phase between the conventional Weyl semimetal and 3D quadrupole topological phases. On the other hand, by regarding the 3D system as $k_z$ dependent 2D systems, the HOWPs serve as the boundary between the QAHI phases and the QTI phase, whereas the conventional Weyl points serve as the boundary between the QAHI phases and the NI phase.

{\it Conclusion and outlook}.---In this Letter, we unveil the concept of HOWPs as a type of Weyl points connected with higher-order topology. When the 3D system is regarded as $k_z$-dependent 2D systems, HOWPs appear as the transition points between the QAHIs and the HOTI with quadrupole topology. Experimental signatures of the higher-order Weyl points consist of the coexistence of Fermi-arc surface states and the topological hinge states, as well as the $k_z$-dependent fractional charge at the hinges. The HOWPs can be realized in tetragonal chiral crystals with the $S_{4z}$, $C_{2x}$, $C_{2y}$, and ${\cal T}$ symmetries. Potential material candidates in electronic and acoustic systems are suggested in the Supplemental Material~\cite{SM}. We remark that due to the richness of HOTIs, there are many other types of HOWPs where the HOTI phase can be realized without the quadrupole topology (e.g., HOTI phases with quantized Wannier centers~\cite{Hughes2019corcharge}). The richness of higher-order topological degeneracies opens a new frontier in topological physics.

\begin{acknowledgments}
This work is supported by the Jiangsu specially-appointed professor funding, the National Natural Science Foundation of China under Grant No 11675116, 11904060, and a project funded by the Priority Academic Program Development of Jiangsu Higher Education Institutions (PAPD).
\end{acknowledgments}

\end{document}